\documentclass{aastex}
\usepackage{spr-astr-addons}
\usepackage{url}
\usepackage{graphicx}
\usepackage{epsfig}
\usepackage{epstopdf}
\usepackage{amsmath}
\RequirePackage{color}
\begin{document}

\title{Gamow-Teller strength distributions and stellar weak-interaction rates for $^{76}$Ge and $^{82}$Se using the deformed pn-QRPA
model}

\shorttitle{Short article title} \shortauthors{Autors et al.}

\author{Jameel-Un Nabi\altaffilmark{1}} \and \author{Mavra Ishfaq\altaffilmark{1}}
\altaffiltext{1}{Faculty of Engineering Sciences,\\GIK Institute of
Engineering Sciences and Technology, Topi 23640, Khyber Pakhtunkhwa,
Pakistan.} \altaffiltext{1}{jameel@giki.edu.pk}
\altaffiltext{2}{mavra.ishfaq34@gmail.com}
\begin{abstract}
We calculate Gamow-Teller strength distributions for $\beta
\beta$-decay nuclei $^{76}$Ge and $^{82}$Se using the deformed
pn-QRPA model. We use a deformed Nilsson basis and consider pairing
correlations within the deformed BCS theory. Ground state
correlations  and two-particle and two-hole mixing states were
included in our pn-QRPA model. Our calculated strength distributions
were compared with experimental data and previous calculation. The
total Gamow-Teller strength and centroid placement calculated in our
model compares well with the measured value.  We calculate
$\beta$-decay and positron capture rates on $^{76}$Ge and $^{82}$Se
in supernovae environments and compare them to those obtained from
experimental data and previous calculation. Our study shows that
positron capture rates command the total weak rates at high stellar
temperatures. We also calculate energy rates of $\beta$-delayed
neutrons and their emission probabilities.

\end{abstract}


\section{Introduction}
Weak rates on nuclei play a crucial role in the presupernova
evolution of massive stars. Their role extends till the star goes
supernova and is intricately connected with the nucleosynthesis
problem. The positron captures and $\beta$-decays play an important
role in the dynamics of the collapsing core in a supernova
\citep{ful82,ful85}. The Gamow Teller (GT) transitions have a vital
and consequential role in various important astrophysical phenomena.
Weak interactions in presupernova stars are known to be dominated by
allowed Fermi (vector-type) and GT (axial-vector-type) transitions.
The calculation of weak interaction rates is very sensitive to the
distribution of the GT$_{\pm}$ strength function. In the GT$_{+}$
strength a proton is changed into a neutron whereas the GT$_{-}$
strength is responsible for transforming a neutron into a proton.
The electron-to baryon ratio $Y_{e }$ reduces due to the electron
capture process and as a result, the nuclear configuration is moved
to more neutron-rich and heavier nuclei \citep{bethe79}. The
first-ever extensive calculation of stellar weak rates was carried
by employing the Independent Particle Model  for iron group nuclei
\citep{ful82,ful85}.  During the supernova and presupernova phases
of a massive star, there is a reasonable probability of occupation
of parent excited states (because of the prevailing high
temperatures and densities) and the total weak interaction rates
have a finite contribution form these excited states. Brink-Axel
hypothesis \citep{Bri55} was adopted by former calculations  of
stellar weak-interaction rates (e.g. \citep{ful82,lan00}) to
approximate the contribution of partial rates from high-lying
excited states. By using this hypothesis it was assumed that the
excited state GT strength distributions are the same as the
calculated ground-state distribution.  However, there is mounting
evidence that the Brink-Axel hypothesis is a poor approximation to
GT strength functions off excited parent states. Nabi and
collaborators
(\cite{Nab05,Nab07,Nab07a,Nab08,Nab09,Nab10,Nab11,Nab12}) calculated
GT strength distributions off parent excited states in a microscopic
fashion for hundreds of nuclei of astrophysical importance and found
out that Brink-Axel hypothesis is a poor approximation for GT off
excited states. \cite{Mis14,Fra97} also advocated for the failure of
Brink-Axel hypothesis in GT transitions. Thus, whenever it is
computationally feasible, one should avoid use of the GT Brink Axel
hypothesis. The proton neutron quasiparticle random phase
approximation (pn-QRPA) theory is an efficient and convenient way to
generate the GT transitions \citep{Vogel86,Engel88}. There are two
main advantages of using the pn-QRPA theory. The first big advantage
is that the pn-QRPA model gets rid of the poor Brink's hypothesis
and calculates ground as well as excited state GT strength
distributions in a microscopic fashion. Secondly the model can carry
reliable calculation of stellar weak rates for any arbitrary heavy
system of nucleons (as calculations are performed in a model space
of up to seven major oscillator shells).

In this paper we perform ground and excited states GT calculation
for two nuclei having energetically allowed $\beta\beta$-channels
available. Only few nuclei in nature undergo allowed
$\beta\beta$-decay which is a two-step second-order weak process.
The intermediate $\beta$-decay state is energetically inaccessible
and is passed through as a virtual intermediate state. Earlier Madey
and collaborators \citep{mad89} studied the excitation-energy
distributions of transition strength to 1$^{+}$ states excited via
the $(p,n)$ reaction at 134.4 MeV on targets of $^{76}$Ge and
$^{82}$Se (along with two heavy isotopes of Te) for excitation
energies up to 25 MeV. A better understanding of the low-lying part
of the GT strength distribution (theoretical as well as
experimental) of $^{76}$Ge and $^{82}$Se is in order to improve the
predictions of $\beta\beta$-decay rates. Recently Ha and Cheoun
\citep{ha15} employed a deformed pn-QRPA model using  the Brueckner
G-matrix based on the CD Bonn potential to calculate ground-state GT
strength distributions of these nuclei. However there was a need to
also calculate excited state GT strength distributions and
associated stellar weak rates for these nuclei. We chose a deformed
pn-QRPA model with schematic separable interaction to calculate
ground- and excited-states GT$_{-}$ transitions, $\beta$-decay $\&$
positron capture rates, energy rates of emitted neutrons from
daughter nuclei and probabilities of $\beta$-delayed neutron
emissions for $^{76}$Ge and $^{82}$Se. All weak interaction rate
calculations were performed in stellar matter.

During the past decades reliable predictions of $\beta$-decay matrix
elements, based on microscopic theories like the pn-QRPA, have
become available \citep{sta90,Hir93}.  From a theoretical point of
view the calculation of additional $\beta$-decay observable like
mean energies provides a valuable test of our understanding of
nuclear structure. Farther away from line of stability, with
decreasing neutron separation energy, $\beta$-delayed neutron
emission occurs as a competitive decay mode to the common $\gamma$
de-excitation. Thus, $\beta$-delayed neutron emission may strongly
affect the prediction of mean $\gamma$ energies. Information on
probabilities of $\beta$-delayed neutron emission ($P_n$) is even
more scarce than that for mean energies. $P_n$ values are also of
interest in astrophysics since they influence the final abundance of
heavy elements synthesized in the r-process. Despite their
importance, $\beta$-delayed neutron emission has often been treated
in rather crude approximations. A reliable prediction of $P_n$
values requires particularly accurate information about the shape of
the $\beta$ strength function, thus underlining the necessity of
microscopic calculations \citep{Hir92}.

This paper is organized in the following format. Section~2 describes
the essential formalism for the calculation of GT strength
distribution and associated stellar weak rates using the pn-QRPA
theory. We present our results in Section~3 where we also compare
them with experimental data and previous calculation. We summarize
the main points and conclude finally in Section~4.

\section{The pn-QRPA formalism}
The current deformed pn-QRPA model differs in two ways from the
deformed pn-QRPA calculation performed by \citep{ha15}. We took a
deformed Nilsson single particle basis whereas Ha $\&$ Cheoun took a
deformed, axially symmetric Woods--Saxon potential. The other main
difference is in choice of interaction.  Ha $\&$ Cheoun used the
Brueckner G-matrix based on the CD Bonn potential whereas we
incorporated a schematic separable interaction
\citep{sta90,mut92,Hir93,Hom96}. The advantage of using these
separable GT forces is that the QRPA matrix equation reduces to an
algebraic equation of fourth order, which is much easier to solve as
compared to full diagonalization of the non-Hermitian matrix of
large dimensionality. Solution of algebraic equation saves order of
magnitude in CPU time and allows a microscopic calculation of GT
strength off parent excited states. In our model, proton-neutron
residual interactions occur as particle-hole (characterized by
interaction constant $\chi$) and particle-particle (characterized by
interaction constant $\kappa$) interactions. The particle-particle
interaction was usually neglected in previous $\beta^{-}$-decay
calculations \citep{sta90,Kru84,Moe90,Ben88}. However it was later
found to be important, specially for the calculation of
$\beta^{+}$-decay \citep{sta90,Hir93,Hir91,Suh88}. The incorporation
of particle-particle force leads to a redistribution of the
calculated $\beta$ strength, which is commonly shifted toward lower
excitation energies \citep{Hir93}.  Other variations of the deformed
pn-QRPA formalism also exists in literature. One such example
includes residual spin-isospin forces in the particle-hole and
particle-particle channels based on a deformed Hartree-Fock
calculation with density-dependent Skyrme forces \citep{Sar03}.

In the present pn-QRPA  formalism \citep{mut92}, we start with a
spherical nucleon basis $(c^{\dagger}_{jm}, c_{jm})$, possessing
total angular momentum j and its z-component $m$. This is later
converted to a deformed (axial-symmetric) basis
$(d^{\dagger}_{m\alpha}, d_{m\alpha})$

\begin{equation}\label{df}
d^{\dagger}_{m\alpha}=\Sigma_{j}D^{m\alpha}_{j}c^{\dagger}_{jm}.
\end{equation}
The transformation matrix $D$ is a set of Nilsson eigenfunctions,
and $\alpha$ (excluding $m$, which specify the Nilsson eigenstates)
represents additional quantum numbers. We performed the BCS
calculation for the neutron and proton systems separately. Here we
assumed a constant pairing force with a force strength G($G_p$ and
$G_n$ for neutrons and protons, respectively),
\begin{equation}\label{pr}
\begin{split}
V_{pair}=-G\sum_{jmj^{'}m^{'}}(-1)^{l+j-m}c^{\dagger}_{jm}c^{\dagger}_{j-m}\\
(-1)^{l^{'}+j^{'}-m^{'}} c_{j^{'}-m}c_{j^{'}m^{'}},
\end{split}
\end{equation}
where the sum over $m$ and $m^{'}$ is constrained to $m$, $m^{'}$
$>$ 0, and $l$ denotes the orbital angular momentum. The BCS
calculation gives occupation amplitudes $u_{m\alpha}$ and
$v_{m\alpha}$ (which satisfy $u^{2}_{m\alpha}$+$v^{2}_{m\alpha}$=1)
and quasiparticle (q.p.) energies $\varepsilon_{m\alpha}$. We later
introduced a q.p. basis $(a^{\dagger}_{m\alpha}, a_{m\alpha})$ by
the distinct Bogoliubov transformation
\begin{equation}\label{qbas}
a^{\dagger}_{m\alpha}=u_{m\alpha}d^{\dagger}_{m\alpha}-v_{m\alpha}d_{\bar{m}\alpha}\\
a^{\dagger}_{\bar{m}\alpha}=u_{m\alpha}d^{\dagger}_{\bar{m}\alpha}+v_{m\alpha}d_{m\alpha},
\end{equation}
where $\bar{m}$ is the time reversed state of $m$ and
$a^{\dagger}/a$ are the q.p. creation/annihilation operators which
enter the RPA equation. Creation operators of QRPA phonons were
defined by
\begin{equation}\label{co}
A^{\dagger}_{\omega}(\mu)=\sum_{pn}[X^{pn}_{\omega}(\mu)a^{\dagger}_{p}a^{\dagger}_{n}-Y^{pn}_{\omega}(\mu)a_{n}a_{\overline{p}}].
\end{equation}
In Eq.~\ref{co}, indices $p$ and $n$ denote $m_{p}\alpha_{p}$ and
$m_{n}\alpha_{n}$, respectively, and differentiate between proton
and neutron single-(quasi)-particle states. The sum runs over
proton-neutron pairs which satisfy $\mu=m_{p}-m_{n}$ and
$\pi_{p}.\pi_{n}$=1, with $\pi$ being parity. We took the $ph$ GT
force in our RPA calculation as
\begin{equation}\label{ph}
V^{ph}= +2\chi\sum^{1}_{\mu= -1}(-1)^{\mu}Y_{\mu}Y^{\dagger}_{-\mu}\\
\end{equation}
\begin{equation}\label{ph}
Y_{\mu}= \sum_{j_{p}m_{p}j_{n}m_{n}}<j_{p}m_{p}\mid
t-\sigma_{\mu}\mid
j_{n}m_{n}>c^{\dagger}_{j_{p}m_{p}}c_{j_{n}m_{n}},
\end{equation}
and the $pp$ GT force as
\begin{equation}\label{ph}
V^{pp}= -2\kappa\sum^{1}_{\mu=
-1}(-1)^{\mu}P_{\mu}P^{\dagger}_{-\mu}.
\end{equation}
Here, the $ph(pp)$ force is defined with the positive(negative)
sign, since the $ph(pp)$ force in $J^{\pi}=1^{+}$ channel is
generally repulsive (attractive), and then the interaction strength
$\chi$ and $\kappa$ take positive values. We took the value of
$\chi$ = 0.01 MeV and $\kappa$ = 0.0955 MeV in our calculation. In
order to reproduce the GT strength the interaction strength
parameters $\chi$  and $\kappa$ were adjusted in our calculation.

In RPA equation matrix elements of the forces are separable,
\begin{equation}\label{phs}
V^{ph}_{pn,p^{'}n^{'}}= +2\chi f_{pn}(\mu)f_{p^{'}n^{'}(\mu)},\\
\end{equation}
\begin{equation}\label{pps}
V^{pp}_{pn,p^{'}n^{'}}= -2\kappa f_{pn}(\mu)f_{p^{'}n^{'}(\mu)},
\end{equation}
with
\begin{equation}\label{f}
f_{pn}(\mu)=\sum_{j_{p}j_{n}}D^{m_{p}\alpha_{p}}_{j_{p}}D^{m_{n}\alpha_{n}}_{j_{n}}<j_{p}m_{p}\mid
t-\sigma_{\mu}\mid j_{n}m_{n}>,
\end{equation}
which are single-particle GT transition amplitudes defined in the
Nilsson basis. For the separable forces, the matrix equation can be
articulated more clearly as
\begin{equation}\label{x}
\begin{split}
X^{pn}_{\omega}=\frac{1}{\omega-\varepsilon_{pn}}[2\chi(q_{pn}Z^{-}_{\omega}+\tilde{q_{pn}}Z^{+}_\omega)\\
-2\kappa(q^{U}_{pn}Z^{- -}_{\omega}+q^{V}_{pn}Z^{+ +}_{\omega})],
\end{split}
\end{equation}
\begin{equation}\label{y}
\begin{split}
Y^{pn}_{\omega}=\frac{1}{\omega+\varepsilon_{pn}}[2\chi(q_{pn}Z^{+}_{\omega}+\tilde{q_{pn}}Z^{-}_\omega)\\
+2\kappa(q^{U}_{pn}Z^{+ +}_{\omega}+q^{V}_{pn}Z^{- -}_{\omega})],
\end{split}
\end{equation}
where $\varepsilon_{pn}=\varepsilon_{p}+\varepsilon_{n}$,\\ \\
$q_{pn}=f_{pn}u_pv_n$, $q_{pn}^{U}=f_{pn}u_pu_n$,\\ \\
$\tilde q_{pn}=f_{pn}v_pu_n$, $q_{pn}^{V}=f_{pn}v_pv_n$,\\ \\
\begin{equation}\label{Z-}
Z^{-}_{\omega}= \sum_{pn}(X^{pn}_{\omega}q_{pn}-Y^{pn}_{\omega}\tilde q_{pn})\\
\end{equation}
\begin{equation}\label{Z+}
Z^{+}_{\omega}= \sum_{pn}(X^{pn}_{\omega}\tilde q_{pn}-Y^{pn}_{\omega}q_{pn})\\
\end{equation}
\begin{equation}\label{Z--}
Z^{- -}_{\omega}= \sum_{pn}(X^{pn}_{\omega}q^{U}_{pn}+Y^{pn}_{\omega}q^{V}_{pn})\\
\end{equation}
\begin{equation}\label{Z++}
Z^{+ +}_{\omega}=
\sum_{pn}(X^{pn}_{\omega}q^{V}_{pn}+Y^{pn}_{\omega}q^{U}_{pn}).
\end{equation}
After some simplification we finally get a matrix equation,
\begin{equation}\label{M}
M z=0,
\end{equation}
which has a solution given by
\begin{equation}\label{DM0}
det M=0,
\end{equation}
thereby reducing the solution of conventional  eigenvalue problem of
the RPA equation to finding roots of algebraic equation
(Eq.~\ref{DM0}). For details of solution we refer to \citep{mut92}.

Using the normalization condition of the phonon amplitudes
\begin{equation}\label{DM1}
\sum_{pn}[(X^{pn}_{\omega})^{2}-(Y^{pn}_{\omega})^{2}]=1,
\end{equation}
the absolute values are determined by inserting $Z_{\omega}'s$ into
Eq.~\ref{x} and Eq.~\ref{y}.

GT transition amplitudes from the QRPA ground state $|-\rangle$
(QRPA vacuum; $A_{\omega}(\mu)|-\rangle=0)$ to one-phonon states
$|\omega,\mu\rangle=A^{\dagger}_{\omega}(\mu)|-\rangle$ are then
calculated as
\begin{equation}\label{DM2}
\langle \omega,\mu|t_{\pm}\sigma_{\mu}|-\rangle=\mp
Z^{\pm}_{\omega}.
\end{equation}
Excitation energies of the one-phonon states are given by
$\omega-(\varepsilon_{p}+\varepsilon_{n})$, where $\varepsilon_{p}$
and $\varepsilon_{n}$ are energies of the single q.p. states of the
smallest q.p. energy in the proton and neutron systems,
respectively.

The RPA is solved for excitations from the $J^{\pi}=0^{+}$ ground
state of an even-even nucleus. In the present model, excited states
of $^{76}$Ge and $^{82}$Se are obtained by one-proton (or
one-neutron) excitations. They are described, in the q.p. picture,
by adding two-proton (two-neutron) q.p.'s to the ground state
\citep{mut92}. Transitions from these initial states are possible to
final proton-neutron q.p. pair states in the odd-odd daughter
nucleus. The transition amplitudes and their reduction to correlated
($c$) one-q.p. states are given by
\begin{eqnarray}
<p^{f}n_{c}^f \mid t_{\pm}\sigma_{-\mu} \mid p_{1}^{i}p_{2c}^{i}> \nonumber \\
 = -\delta (p^{f},p_{2}^{i}) <n_{c}^{f} \mid t_{\pm}\sigma_{-\mu} \mid
 p_{1c}^{i}> \nonumber \\
+\delta (p^{f},p_{1}^{i}) <n_{c}^{f} \mid t_{\pm}\sigma_{-\mu} \mid
p_{2c}^{i}>
 \label{first}
\end{eqnarray}
\begin{eqnarray}
\begin{split}
<p^{f}n_{c}^f \mid t_{\pm}\sigma_{\mu} \mid n_{1}^{i}n_{2c}^{i}> \nonumber \\
 = +\delta (n^{f},n_{2}^{i}) <p_{c}^{f} \mid t_{\pm}\sigma_{\mu} \mid
 n_{1c}^{i}> \nonumber \\
-\delta (n^{f},n_{1}^{i}) <p_{c}^{f} \mid t_{\pm}\sigma_{\mu} \mid
n_{2c}^{i}>
\end{split}
\end{eqnarray}
where $\mu$ = -1, 0, 1, are the spherical components of the spin
operator.

For the odd-odd daughter nucleus the ground state is assumed to be a
proton-neutron q.p. pair state of smallest energy. Excited states
are expressed in q.p. transformation by two-q.p. states
(proton-neutron pair states) or by four-q.p. states (two-proton or
two-neutron q.p. states). Reduction of two-q.p. states into
correlated ($c$) one-q.p. states is given as
\begin{eqnarray}
<p_{1}^{f}p_{2c}^{f} \mid t_{\pm}\sigma_{\mu} \mid p^{i}n_{c}^{i}> \nonumber\\
= \delta(p_{1}^{f},p^{i}) <p_{2c}^{f} \mid t_{\pm}\sigma_{\mu} \mid
n_{c}^{i}> \nonumber\\ - \delta(p_{2}^{f},p^{i}) <p_{1c}^{f} \mid
t_{\pm}\sigma_{\mu} \mid n_{c}^{i}>
\end{eqnarray}
\begin{eqnarray}
<n_{1}^{f}n_{2c}^{f} \mid t_{\pm}\sigma_{-\mu} \mid p^{i}n_{c}^{i}> \nonumber\\
= \delta(n_{2}^{f},n^{i}) <n_{1c}^{f} \mid t_{\pm}\sigma_{-\mu} \mid
p_{c}^{i}> \nonumber\\ - \delta(n_{1}^{f},n^{i}) <n_{2c}^{f} \mid
t_{\pm}\sigma_{-\mu} \mid p_{c}^{i}>
\end{eqnarray}
while the four-q.p. states are simplified as
\begin{eqnarray}
<p_{1}^{f}p_{2}^{f}n_{1}^{f}n_{2c}^{f} \mid t_{\pm}\sigma_{-\mu}
\mid p_{1}^{i}p_{2}^{i}p_{3}^{i}n_{1c}^{i}>
\nonumber\\
=\delta (n_{2}^{f},n_{1}^{i})[ \delta (p_{1}^{f},p_{2}^{i})\delta
(p_{2}^{f},p_{3}^{i})\nonumber\\
<n_{1c}^{f} \mid t_{\pm}\sigma_{-\mu} \mid p_{1c}^{i}> \nonumber\\
-\delta (p_{1}^{f},p_{1}^{i}) \delta (p_{2}^{f},p_{3}^{i})
<n_{1c}^{f} \mid t_{\pm}\sigma_{-\mu} \mid p_{2c}^{i}>
\nonumber\\
+\delta(p_{1}^{f},p_{1}^{i}) \delta (p_{2}^{f},p_{2}^{i}) \nonumber\\
<n_{1c}^{f} \mid t_{\pm}\sigma_{-\mu} \mid p_{3c}^{i}>]\nonumber\\
-\delta (n_{1}^{f},n_{1}^{i})[ \delta (p_{1}^{f},p_{2}^{i})\delta
(p_{2}^{f},p_{3}^{i})\nonumber\\
<n_{2c}^{f} \mid t_{\pm}\sigma_{-\mu} \mid p_{1c}^{i}> \nonumber\\
-\delta (p_{1}^{f},p_{1}^{i}) \delta (p_{2}^{f},p_{3}^{i})
<n_{2c}^{f} \mid t_{\pm}\sigma_{-\mu} \mid p_{2c}^{i}>
\nonumber\\+\delta
(p_{1}^{f},p_{1}^{i}) \delta (p_{2}^{f},p_{2}^{i})\nonumber\\
<n_{2c}^{f}\mid t_{\pm}\sigma_{-\mu} \mid p_{3c}^{i}>]
\end{eqnarray}
\begin{eqnarray}
<p_{1}^{f}p_{2}^{f}p_{3}^{f}p_{4c}^{f} \mid t_{\pm}\sigma_{\mu} \mid
p_{1}^{i}p_{2}^{i}p_{3}^{i}n_{1c}^{i}>
\nonumber\\
=-\delta (p_{2}^{f},p_{1}^{i}) \delta (p_{3}^{f},p_{2}^{i})\delta
(p_{4}^{f},p_{3}^{i})
<p_{1c}^{f} \mid t_{\pm}\sigma_{\mu} \mid n_{1c}^{i}> \nonumber\\
+\delta (p_{1}^{f},p_{1}^{i}) \delta (p_{3}^{f},p_{2}^{i}) \delta
(p_{4}^{f},p_{3}^{i})
<p_{2c}^{f} \mid t_{\pm}\sigma_{\mu} \mid n_{1c}^{i}> \nonumber\\
-\delta (p_{1}^{f},p_{1}^{i}) \delta (p_{2}^{f},p_{2}^{i}) \delta
(p_{4}^{f},p_{3}^{i})
<p_{3c}^{f} \mid t_{\pm}\sigma_{\mu} \mid n_{1c}^{i}> \nonumber\\
+\delta (p_{1}^{f},p_{1}^{i}) \delta (p_{2}^{f},p_{2}^{i}) \delta
(p_{3}^{f},p_{3}^{i}) <p_{4c}^{f} \mid t_{\pm}\sigma_{\mu} \mid
n_{1c}^{i}>
\end{eqnarray}
\begin{eqnarray}
<p_{1}^{f}p_{2}^{f}n_{1}^{f}n_{2c}^{f} \mid t_{\pm}\sigma_{\mu} \mid
p_{1}^{i}n_{1}^{i}n_{2}^{i}n_{3c}^{i}>
\nonumber\\
=\delta (p_{1}^{f},p_{1}^{i})[ \delta (n_{1}^{f},n_{2}^{i})\delta
(n_{2}^{f},n_{3}^{i})\nonumber\\
<p_{2c}^{f} \mid t_{\pm}\sigma_{\mu} \mid n_{1c}^{i}> \nonumber\\
-\delta (n_{1}^{f},n_{1}^{i}) \delta (n_{2}^{f},n_{3}^{i})
<p_{2c}^{f} \mid t_{\pm}\sigma_{\mu} \mid n_{2c}^{i}>\nonumber\\
+\delta
(n_{1}^{f},n_{1}^{i}) \delta (n_{2}^{f},n_{2}^{i})\nonumber\\
<p_{2c}^{f} \mid t_{\pm}\sigma_{\mu} \mid n_{3c}^{i}>]\nonumber\\
-\delta (p_{2}^{f},p_{1}^{i})[ \delta (n_{1}^{f},n_{2}^{i})\delta
(n_{2}^{f},n_{3}^{i})\nonumber\\
<p_{1c}^{f} \mid t_{\pm}\sigma_{\mu} \mid n_{1c}^{i}> \nonumber\\
-\delta (n_{1}^{f},n_{1}^{i}) \delta (n_{2}^{f},n_{3}^{i})
<p_{1c}^{f} \mid t_{\pm}\sigma_{\mu} \mid n_{2c}^{i}>\nonumber\\
+\delta
(n_{1}^{f},n_{1}^{i}) \delta (n_{2}^{f},n_{2}^{i}) \nonumber\\
<p_{1c}^{f} \mid t_{\pm}\sigma_{\mu} \mid n_{3c}^{i}>]
\end{eqnarray}
\begin{eqnarray}
<n_{1}^{f}n_{2}^{f}n_{3}^{f}n_{4c}^{f} \mid t_{\pm}\sigma_{-\mu}
\mid p_{1}^{i}n_{1}^{i}n_{2}^{i}n_{3c}^{i}>
\nonumber\\
= +\delta (n_{2}^{f},n_{1}^{i}) \delta (n_{3}^{f},n_{2}^{i})\delta
(n_{4}^{f},n_{3}^{i})
<n_{1c}^{f} \mid t_{\pm}\sigma_{-\mu} \mid p_{1c}^{i}> \nonumber\\
-\delta (n_{1}^{f},n_{1}^{i}) \delta (n_{3}^{f},n_{2}^{i}) \delta
(n_{4}^{f},n_{3}^{i})
<n_{2c}^{f} \mid t_{\pm}\sigma_{-\mu} \mid p_{1c}^{i}> \nonumber\\
+\delta (n_{1}^{f},n_{1}^{i}) \delta (n_{2}^{f},n_{2}^{i}) \delta
(n_{4}^{f},n_{3}^{i})
<n_{3c}^{f} \mid t_{\pm}\sigma_{-\mu} \mid p_{1c}^{i}> \nonumber\\
-\delta (n_{1}^{f},n_{1}^{i}) \delta (n_{2}^{f},n_{2}^{i}) \delta
(n_{3}^{f},n_{3}^{i}) <n_{4c}^{f} \mid t_{\pm}\sigma_{-\mu} \mid
p_{1c}^{i}> .\label{last}
\end{eqnarray}
For all the given q.p. transition amplitudes [Eqs. ~(\ref{first})-
~(\ref{last})],
the antisymmetrization of the single- q.p. states was taken into account:\\
$ p_{1}^{f}<p_{2}^{f}<p_{3}^{f}<p_{4}^{f}$,\\
$ n_{1}^{f}<n_{2}^{f}<n_{3}^{f}<n_{4}^{f}$,\\
$ p_{1}^{i}<p_{2}^{i}<p_{3}^{i}<p_{4}^{i}$,\\
$ n_{1}^{i}<n_{2}^{i}<n_{3}^{i}<n_{4}^{i}$.\\
GT transitions of phonon excitations for every excited state were
also taken into account. Here it was assumed that the quasiparticles
in the parent nucleus remained in the same quasiparticle orbits. A
detailed description of the formalism can be found in Ref.
\citep{mut92}.

The deformation parameter ($\beta$) was argued to be one of the most
important parameters in pn-QRPA calculation \citep{Ste04} and as
such rather than using $\beta$ values calculated from some
theoretical mass model (as used in earlier calculations of pn-QRPA
capture rates) the experimentally adopted value of $\beta$ for
$^{76}$Ge (0.26) and $^{82}$Se (0.19), extracted by relating the
measured energy of the first $2^{+}$ excited state with the
quadrupole deformation, was taken from \citep{Ram87}. The
incorporation of experimental deformations  led to an overall
improvement in the calculation as discussed earlier in
\citep{Nab09}. Q-values were taken from the mass compilation of
\citep{Aud12}.

For the parent nuclei we considered a maximum of 70,000 states. We
grouped 200 such states in a band of width 50 keV and then each band
was treated as a state in the calculation of weak rates. Within each
band, the GT strength of each contributing state was taken into
account. Increasing further the number of excited states led to an
increase in CPU time up to a factor of ten or more which was not
practical. However, we did assure that a minimum of 10 MeV be
considered for the excitation energy in the parent nucleus. Due to
selection rules for GT transitions in the pn-QRPA model, it was
possible to consider states up to 30 MeV in daughter nuclei. We
multiplied results of pn-QRPA calculated strength by a quenching
factor of $f_{q}^{2}$ = (0.55)$^{2}$ \citep{nab15b} in order to
compare them with experimental data and prior calculations, and to
later use them in astrophysical reaction rates.

The weak decay rate from the $\mathit{i}$th state of the parent to
the $\mathit{j}$th state of the daughter nucleus is given by

\begin{equation}
\lambda_{ij}^{bd(pc)} =ln2
\frac{f_{ij}^{bd(pc)}(T,\rho,E_{f})}{(ft)_{ij}^{bd(pc)}},
\end{equation}
where $(ft)_{ij}^{bd(pc)}$ is related to the reduced transition
probability $B_{ij}$ of the nuclear transition by

\begin{equation}\label{ft}
(ft)_{ij}^{bd(pc)}=D/B_{ij}.
\end{equation}
The D appearing in Eq.~\ref{ft} is a compound expression of physical
constants,
\begin{equation}
D=\frac{2ln2\hbar^{7}\pi^{3}}{g_{V}^{2}m_{e}^{5}c^{4}},
\end{equation}
and the value of D was taken to be 6146 $\pm$ 6 s adopted from
\citep{Jok02},

\begin{equation}
B_{ij}=B(F)_{ij}+(g_{A}/g_{V})^2 B(GT)_{ij},
\end{equation}
where B(F) and B(GT) are reduced transition probabilities of the
Fermi and ~GT transitions, respectively.

The phase space integral $(f_{ij})^{bd(pc)}$ is an integral over
total energy,
\begin{equation}\label{ps}
f_{ij} = \int_{1}^{w_{m}} w \sqrt{w^{2}-1} (w_{m}-w)^{2} F(+ Z,w)
(1-G_-) dw,
\end{equation}
for electron emission
\begin{equation}\label{pc}
f_{ij}^{bd(pc)} = \int_{w_{l}}^{\infty} w \sqrt{w^{2}-1}
(w_{m}+w)^{2} F(- Z,w) G_+ dw,
\end{equation}
for continuum positron capture (from here onwards we use natural
units, $\hbar=m_{e}=c=1$).

The $G_{+}$ and $G_{-}$ are the positron and electron distribution
functions, respectively. In Eqs.~\ref{ps} and ~\ref{pc}, $w$ is the
total kinetic energy of the electron including its rest mass,
$w_{l}$ is the total capture threshold energy (rest+kinetic) for
positron capture. F($\pm$ Z,w) is the Fermi function and was
calculated according to the procedure adopted by Gove and Martin
\citep{Gov71}. One should note that if the corresponding electron
emission total energy, $w_{m}$, is greater than -1, then $w_{l}=1$,
and if it is less than or equal to 1, then $w_{l}=\mid w_{m} \mid$.
$w_{m}$ is the total $\beta$-decay energy,
\begin{equation}
w_{m} = m_{p}-m_{d}+E_{i}-E_{j},
\end{equation}
where m$_{p}$ and E$_{i}$ are mass and excitation energies of the
parent nucleus, and m$_{d}$ and E$_{j}$ of the daughter nucleus,
respectively.

The number density of protons associated with electrons and nuclei
is $\rho Y_{e} N_{A}$, where $\rho$ is the baryon density, $Y_{e}$
is the ratio of electron number to the baryon number, and $N_{A}$ is
the Avogadro number.
\begin{equation}\label{ye}
\rho Y_{e} = \frac{1}{\pi^{2}N_{A}}(\frac {m_{e}c}{\hbar})^{3}
\int_{0}^{\infty} (G_{-}-G_{+}) p^{2}dp,
\end{equation}
where $p=(w^{2}-1)^{1/2}$ is lepton's momentum, and Eq.~\ref{ye} has
units of \textit{moles $cm^{-3}$}. This equation was used for an
iterative calculation of Fermi energies for selected values of $\rho
Y_{e}$ and $T$.

The total capture/$\beta^{-}$-decay rate per unit time per nucleus
is finally given by
\begin{equation}
\lambda^{bd(pc)} =\sum _{ij}P_{i} \lambda _{ij}^{bd(pc)}.
\label{total rate}
\end{equation}
The summation over the initial and final states was carried out
until satisfactory convergence was achieved in stellar weak rates
calculation. Here $P_{i} $ is the probability of occupation of
parent excited states and follows the normal Boltzmann distribution.

It was assumed in our calculation that all daughter excited states,
with energy greater than the separation energy of neutrons ($S_{n}$)
decay by emission of neutrons. The neutron energy rate from the
daughter nucleus was calculated using
\begin{equation}\label{ln}
\lambda^{n} = \sum_{ij}P_{i}\lambda_{ij}(E_{j}-S_{n}),
\end{equation}
for all $E_{j} > S_{n}$.

The probability of $\beta$-delayed neutron emission was calculated
by
\begin{equation}\label{pn}
P^{n} =
\frac{\sum_{ij\prime}P_{i}\lambda_{ij\prime}}{\sum_{ij}P_{i}\lambda_{ij}},
\end{equation}
where $j\prime$ are states in the daughter nucleus for which
$E_{j\prime} > S_{n}$. In Eqs.~(\ref{ln} and \ref{pn}),
$\lambda_{ij(\prime)}$ is the sum of the positron capture and
electron decay rates, for the transition $i$ $\rightarrow$
$j(j\prime)$.

\section{Results and Discussions}
\begin{figure}
\includegraphics[scale=0.5]{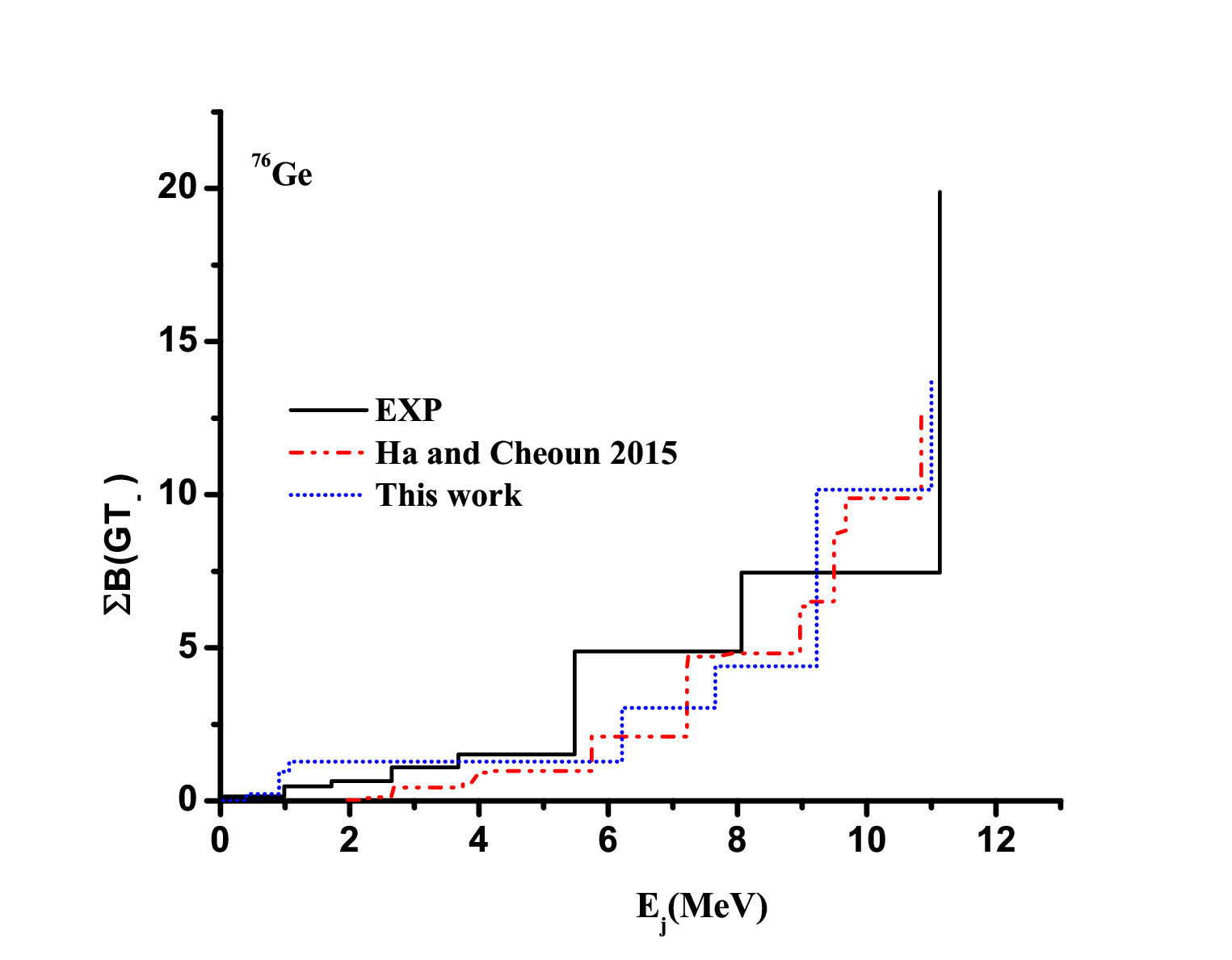}
\normalsize \caption{Our calculated cumulative B(GT$_{-}$)
distribution for the ground state of $^{76}$Ge compared with
experimental data \citep{mad89} and the Ha and Cheoun \citep{ha15}
calculation.}\label{gt76}
\end{figure}

The cumulative GT$_{-}$ strength for ground state of $^{76}$Ge is
shown in Fig.~\ref{gt76}. Here the abscissa represents daughter
excitation energy in units of MeV. We show our calculated GT$_{-}$
strength in the energy interval 0 -- 12 MeV. We also show the GT
strength calculation of Ha $\&$ Cheoun \citep{ha15} in
Fig.~\ref{gt76}.  Excitation-energy distributions of transition
strength to 1$^{+}$ states  at 134.4 MeV on various targets
(including $^{76}$Ge and $^{82}$Se) were measured for excitation
energies up to 25 MeV via the $(p, n)$ reaction by Madey and
collaborators \citep{mad89}. The measured data is also shown in
Fig.~\ref{gt76}.  It is clear from Fig.~\ref{gt76} that the two
calculations are in decent agreement with the measured total
strength. Table~\ref{cent table} shows that the pn-QRPA model places
the GT centroid at 8.66 MeV energy  in daughter which is in good
agreement with the measured centroid energy placed at 9.10 MeV. The
Ha $\&$ Cheoun calculation leads to a much higher centroid. The
pn-QRPA calculated total GT$_{-}$ strength of 16.30 is also in
better comparison with measured strength value of 19.89. Ha $\&$
Cheoun calculated strength is also higher. It is to be noted that
the Ha $\&$ Cheoun calculation did not employ any quenching factor
in their calculation.

\begin{figure}
\includegraphics[scale=0.5]{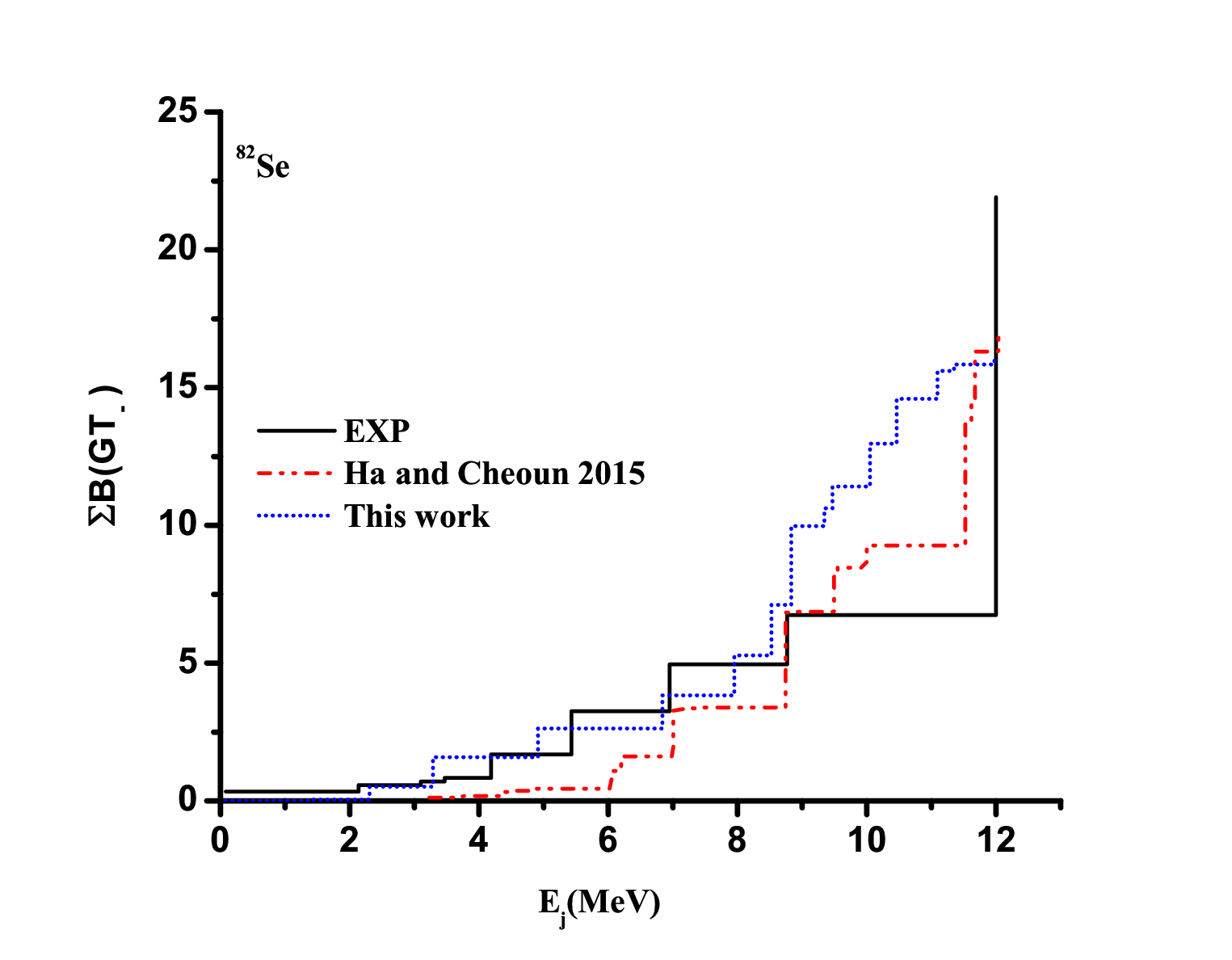}
\normalsize \caption{Same as Fig.~1 but for $^{82}$Se.}\label{gt82}
\end{figure}

Similarly Fig.~\ref{gt82} displays calculated and measured GT$_{-}$
strength for ground state of $^{82}$Se. Measured data was taken from
the same $(p, n)$  experiment \citep{mad89}. The experimental
GT$_{-}$ strength is observed  between 0 -- 12 MeV in daughter. The
Ha $\&$ Cheoun model calculates strength in the range 3 -- 19 MeV.
Our model calculates GT strength in $^{82}$Br between 0 -- 12 MeV as
also supported by measurements \citep{mad89}. Table~\ref{cent table}
once again shows that our calculated GT$_{-}$ strength and centroid
placement agrees well with the measured data. After achieving decent
agreement with experimental data we next proceed to calculate
weak-interaction mediated rates due to $^{76}$Ge and $^{82}$Se  in
stellar environment.

\begin{figure}
\includegraphics[scale=0.5]{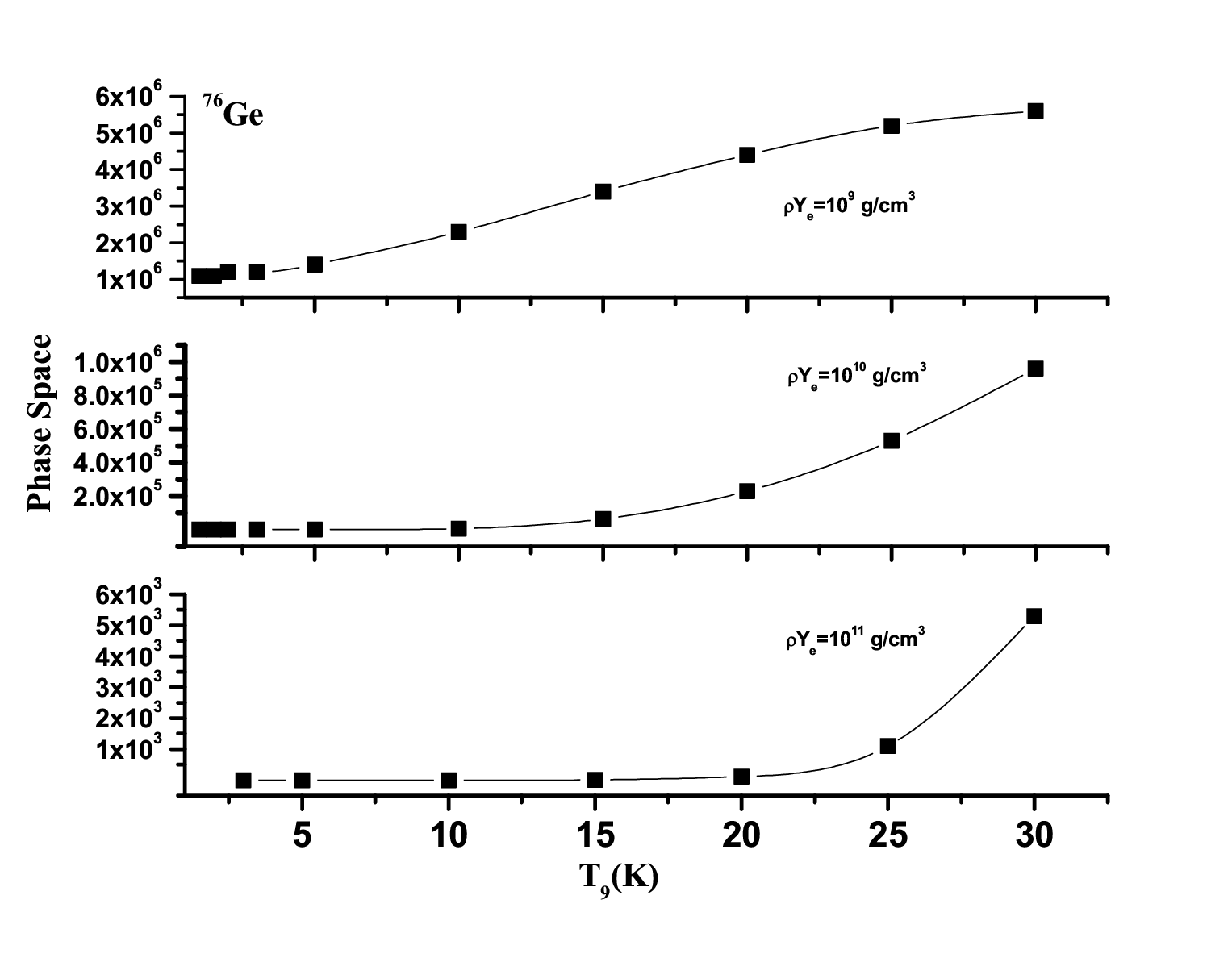}
\normalsize \caption{Calculated phase space for $^{76}$Ge as a
function of stellar temperature and density.}\label{ps76}
\end{figure}

\begin{figure}
\includegraphics[scale=0.5]{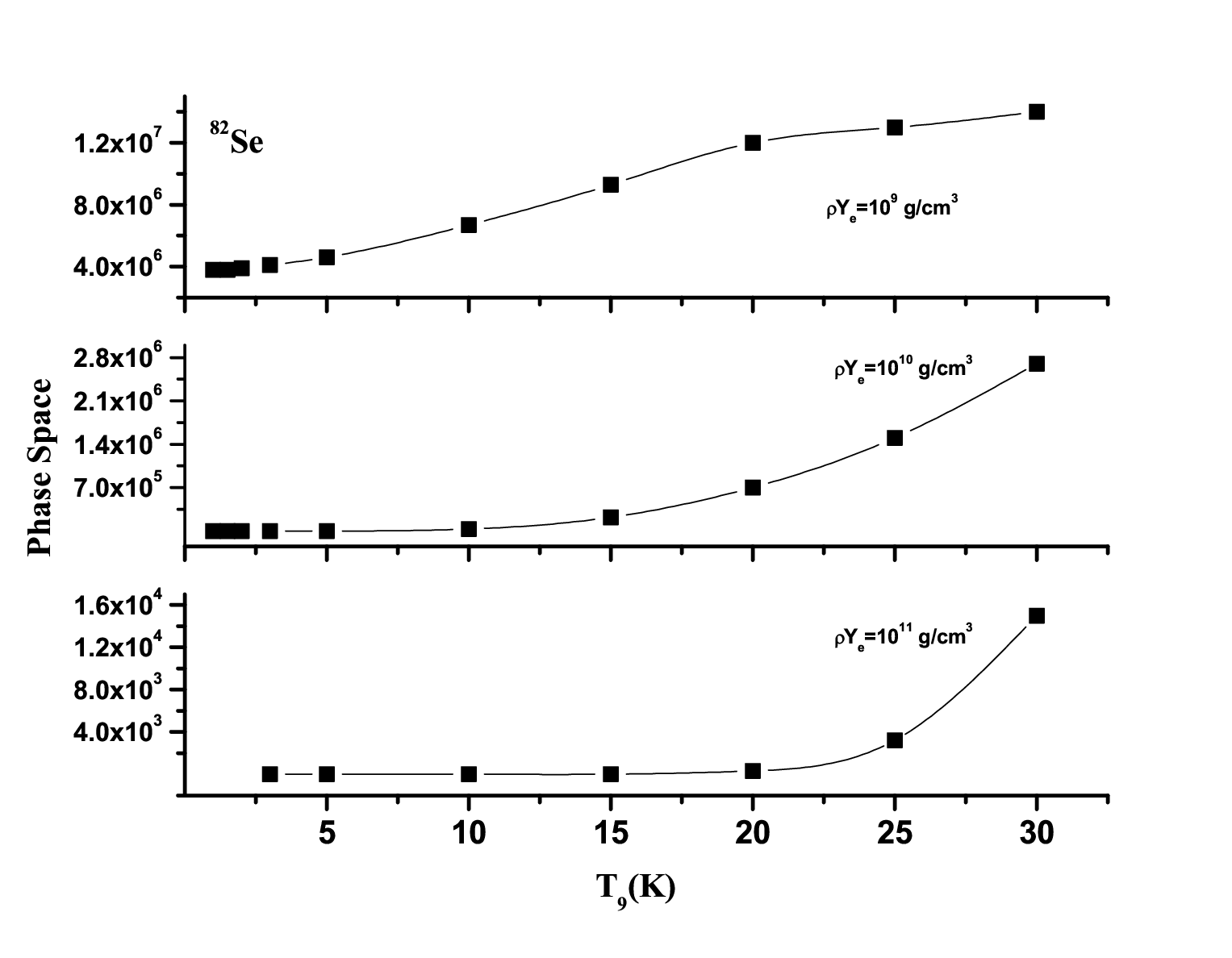}
\normalsize \caption{Same as Fig.~\ref{ps76} but for
$^{82}$Se.}\label{ps82}
\end{figure}

The phase space calculation for $^{76}$Ge, as a function of stellar
temperature and density, is shown in Fig.~\ref{ps76}. The phase
space factors are shown at high stellar density values of 10$^{9}$
g/cm$^{3}$, 10$^{10}$ g/cm$^{3}$ and 10$^{11}$ g/cm$^{3}$  and
stellar temperature T$_{9}$ = 1 - 30 (T$_{9}$ gives the stellar
temperature in units of $10^{9}$ K).  The phase space is essentially
zero at small stellar temperature T$_{9} \sim$ 0.01 and increases
with increasing temperature.  It can be further noted from
Fig.~\ref{ps76} that for stellar density 10$^{10}$ (10$^{11}$)
g/cm$^{3}$  the phase space increases by 7 (26) orders of magnitude
as the stellar temperature goes from T$_{9}$ = 3 to 30.  It is seen
from Fig.~\ref{ps76} that for density 10$^{9}$ g/cm$^{3}$ the phase
space factor increases almost linearly as T$_{9}$ = 1 to 30. At
higher densities the increment is exponential. At a fixed stellar
temperature, the phase space remains the same as the core stiffens
from 10 to 10$^{7}$ g/cm$^{3}$ stellar density. This is because the
lepton distribution function at a fixed temperature changes
appreciably only once the stellar density exceeds 10$^{7}$
g/cm$^{3}$. This happens because of an appreciable increase in the
calculated Fermi energy of the leptons once the stellar density
reaches 10$^{7}$ g/cm$^{3}$ and beyond.  As the stellar core becomes
more and more dense the phase space decreases. The calculated phase
space factor for $^{82}$Se is shown in Fig.~\ref{ps82} and shows a
similar behavior with an order of magnitude bigger phase space
factors at high stellar temperatures.

\begin{figure}
\includegraphics[scale=0.5]{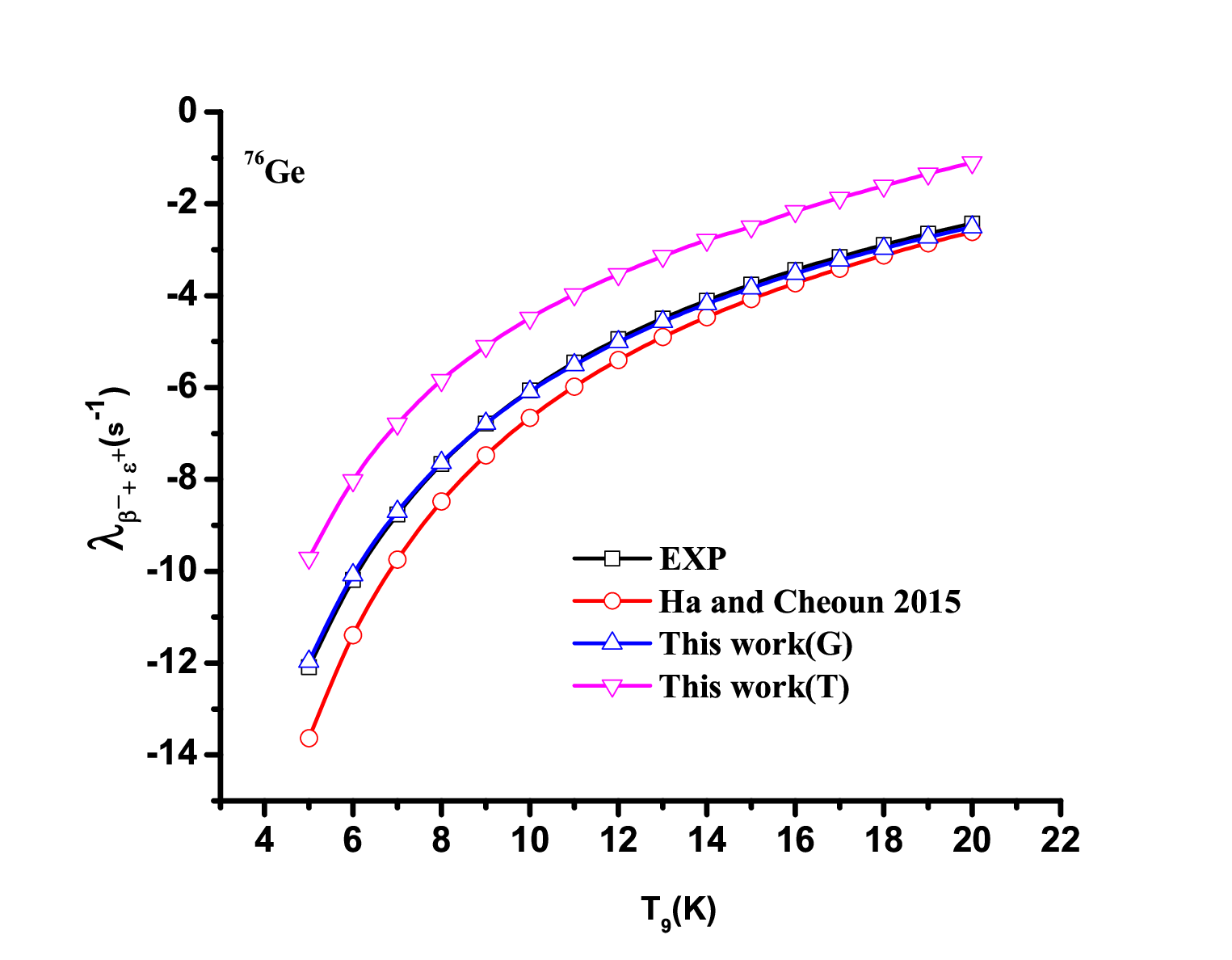}
\normalsize \caption{Our calculated $\beta$-decay and positron
capture rates for $^{76}$Ge compared with other calculations as a
function of stellar temperatures for a fixed stellar density $\rho$
= $10^{9.6}$ g $cm^{-3}$. For explanation of legends see
text.}\label{pc76}
\end{figure}

\begin{figure}
\includegraphics[scale=0.5]{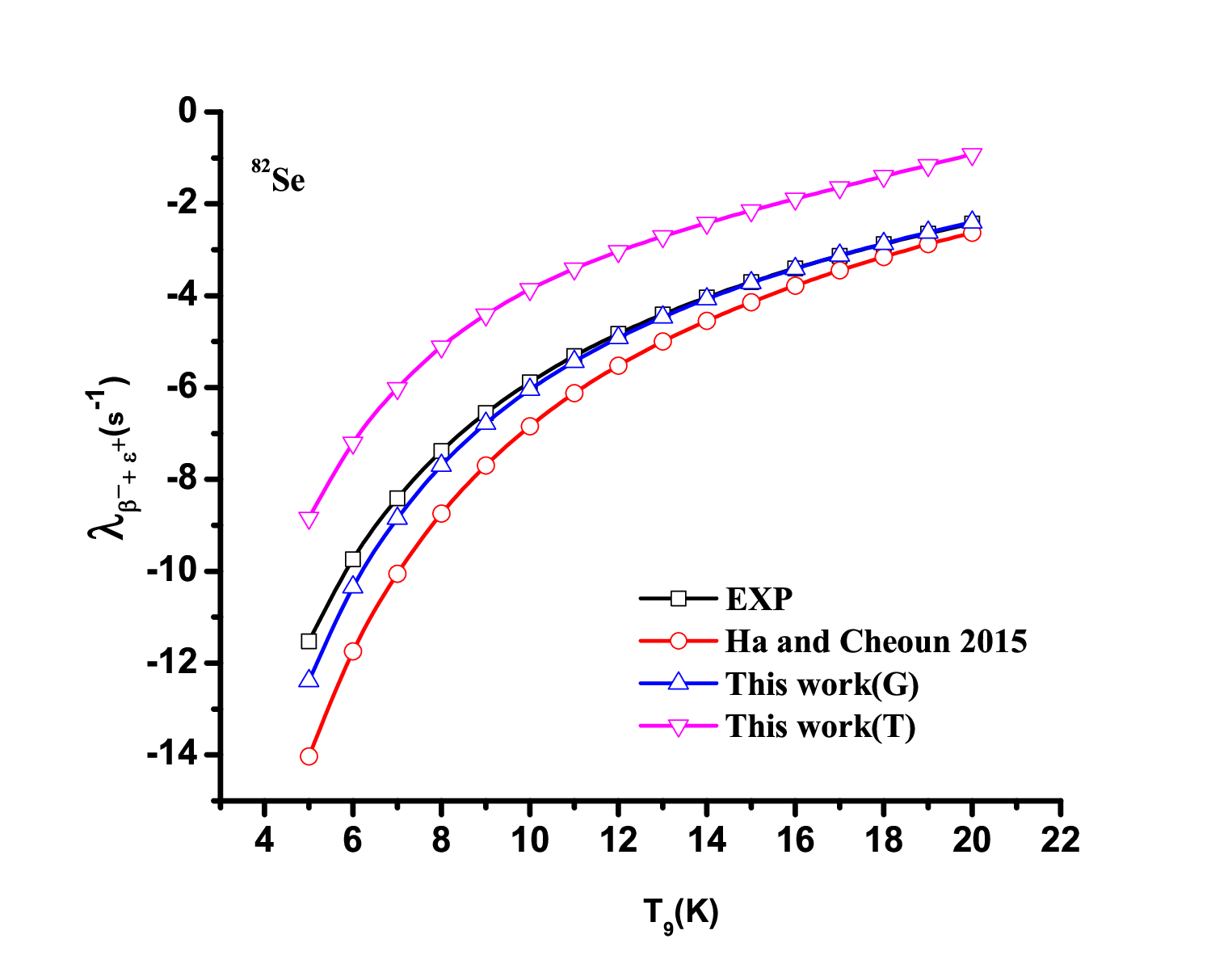}
\normalsize \caption{Same as Fig.~\ref{pc76} but for
$^{82}$Se.}\label{pc82}
\end{figure}

Fig.~\ref{pc76} and Fig.~\ref{pc82} show our (pn-QRPA) calculated
sum of stellar $\beta$-decay and positron capture rates as a
function of stellar temperature at a fixed stellar density of
$10^{9.6}$ g $cm^{-3}$ for $^{76}$Ge and $^{82}$Se, respectively.
The ordinate shows weak rates in logarithmic scale (to base 10) in
units of $s^{-1}$. In these graphs we show our calculated rates with
contribution only from parent ground state (shown as (G)) and those
with contributions from all 200 parent excited states (shown as
(T)). We calculated the weak rates from the the measured GT
distribution \citep{mad89} and the calculated GT distribution of Ha
$\&$ Cheoun \citep{ha15}. The later two rates were calculated on the
basis of ground state GT$_{-}$ strength distributions alone. In
other words for "This work(G)", "Ha and Cheoun 2015" and "EXP" rates
shown in the figures, all parent excited state GT strength functions
were taken to be zero (i.e. in Eq.~\ref{total rate} summation was
performed only on the parent ground state). It can be seen from
Fig.~\ref{pc76} that our rates with only ground state contribution
are in excellent agreement with measured data rates. This we
attribute to the decent comparison of our calculated GT distribution
with the measured GT distribution (see Fig.~\ref{gt76}). The Ha $\&$
Cheoun  model calculated rates are smaller than the rates calculated
using the measured data. This is primarily because of the higher
centroid placement in their model. Our calculated stellar rates,
with contributions from all 200 parent excited states, are enhanced
by orders of magnitude as compared to other three rates because of
finite contribution to the total weak rates from excited state GT
strength distributions. A more or less similar behavior is witnessed
for the case of $^{82}$Se in Fig.~\ref{pc82}.

\begin{figure}
\includegraphics[scale=0.5]{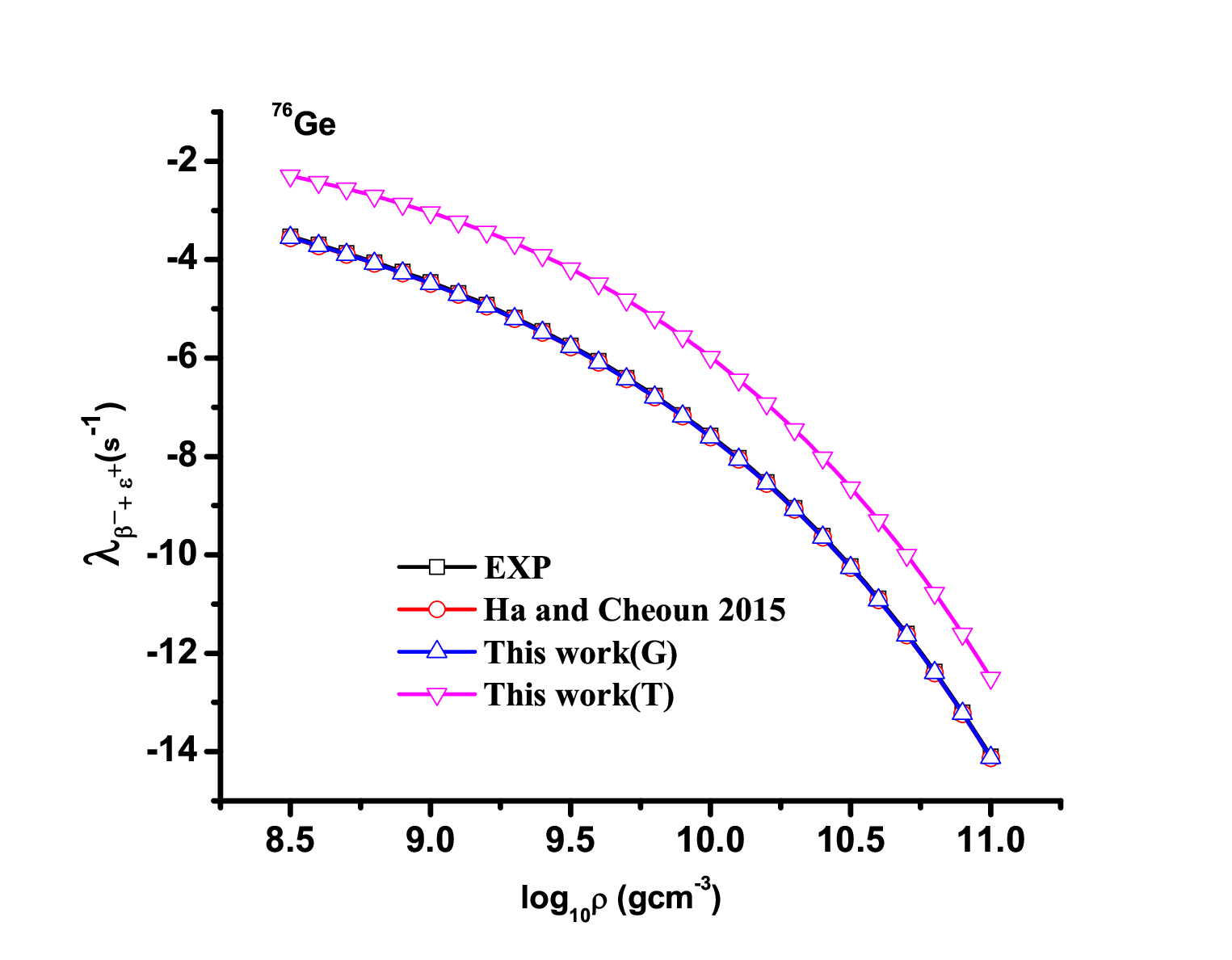}
\normalsize \caption{Same as Fig.~\ref{pc76} but as a function of
stellar density at a fixed temperature of T$_{9}$(K) =
10.}\label{t976}
\end{figure}

\begin{figure}
\includegraphics[scale=0.5]{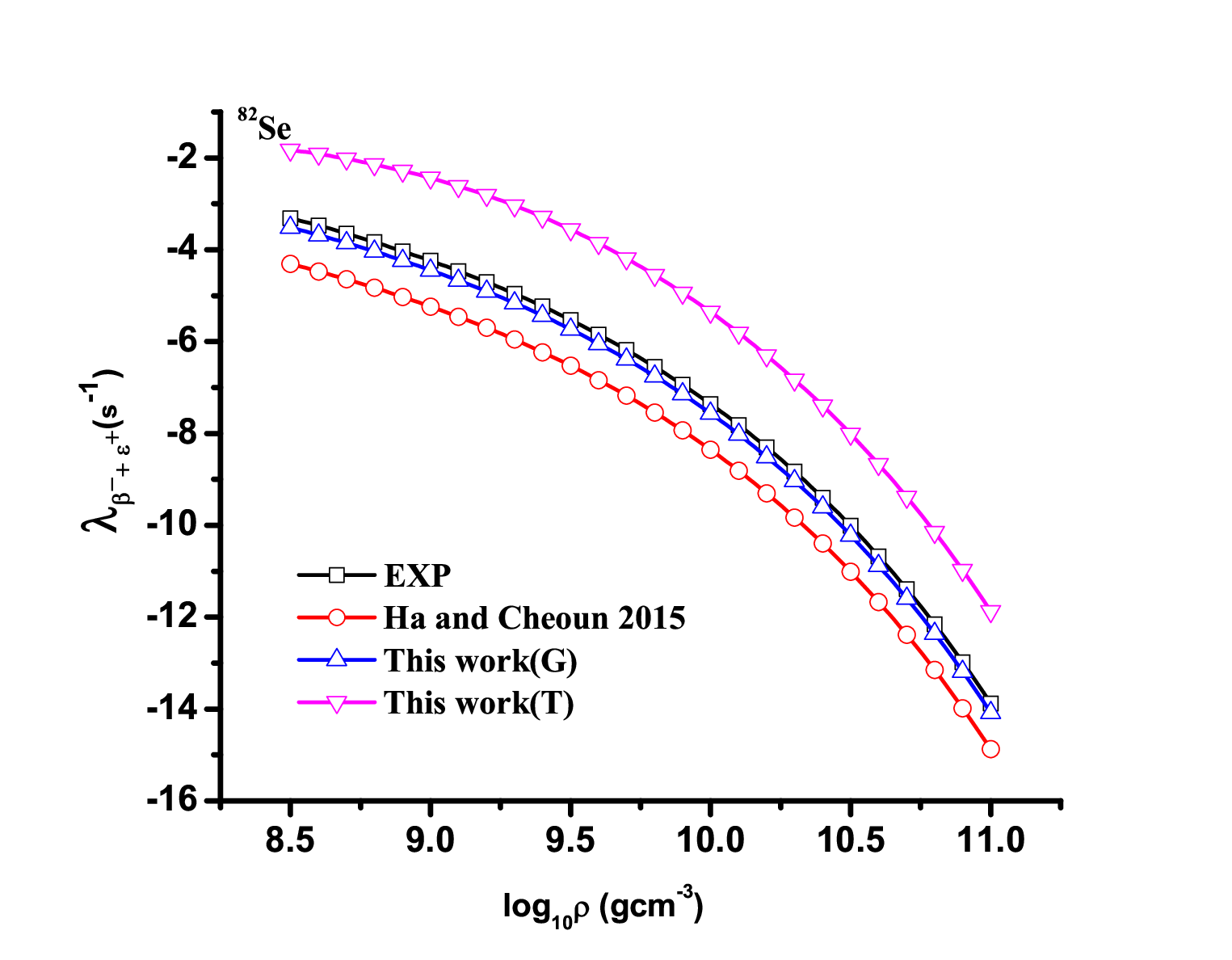}
\normalsize \caption{Same as Fig.~\ref{t976} but for
$^{82}$Se.}\label{t982}
\end{figure}

After showing dependence of calculated weak rates on stellar
temperature, we next show their dependence on stellar density.
Fig.~\ref{t976} and Fig.~\ref{t982} show the calculated rates at a
fixed stellar temperature of T$_{9}$ =10 as the stellar core
stiffens from 10$^{8.5}$ -- 10$^{11}$ g $cm^{-3}$ for the nuclei
$^{76}$Ge and $^{82}$Se, respectively. Fig.~\ref{t976} shows that
the Ha $\&$ Cheoun  and our calculated rates (with only ground state
contribution)  are in excellent agreement with the experimental
rates. This is because at high densities and temperatures, the
electron chemical potential appreciably exceeds the Q value, and in
this phase, weak rates are largely dictated by the total GT strength
and its centroid energy (the low-lying individual transitions does
not matter as much). Table~\ref{cent table} shows a decent
comparison of calculated and measured data for $^{76}$Ge. Once again
our calculated rates with contributions form all parent excited
states are bigger because of finite contribution of excited state GT
strength distributions (missing in the experimental data and other
calculations).

Fig.~\ref{t982} depicts the density dependence of calculated rates
for the case of $^{82}$Se. Here we see that the  calculated rates
for $^{82}$Se show a similar behavior as for the case of $^{76}$Ge.
The Ha $\&$ Cheoun calculated rates are smaller compared to the
experimental rates. Ha $\&$ Cheoun  placed the centroid at roughly
1.5 MeV higher than the centroid placement of the measured data
which resulted in decrement of their calculated rates.

Table~\ref{ratio table} shows the relative contribution of our
calculated stellar positron capture and $\beta$-decay rates to the
total calculated rates for the two nuclei. For the case of $^{76}$Ge
the calculated $\beta$-decay rates are 2 -- 4 orders of magnitude
bigger than the competing positron capture rates at low stellar
temperatures (T$_{9} \leq $ 10). The  positron capture rates compete
well with the $\beta$-decay rates at T$_{9}$ =10. The  positron
capture rates are bigger by one (two) order(s) of magnitude at
T$_{9}$ =20 (T$_{9}$ =30). For the case of $^{82}$Se once again the
$\beta$-decay rates command the total rates at low stellar
temperatures. The positron capture rates are 1 --2 orders of
magnitude bigger at higher  T$_{9}$ values. At high temperatures($kT
>$ 1 MeV), positrons appear via electron-positron pair creation and
their capture rates exceed the competing $\beta$-decay rates.

\begin{figure}
\includegraphics[scale=0.5]{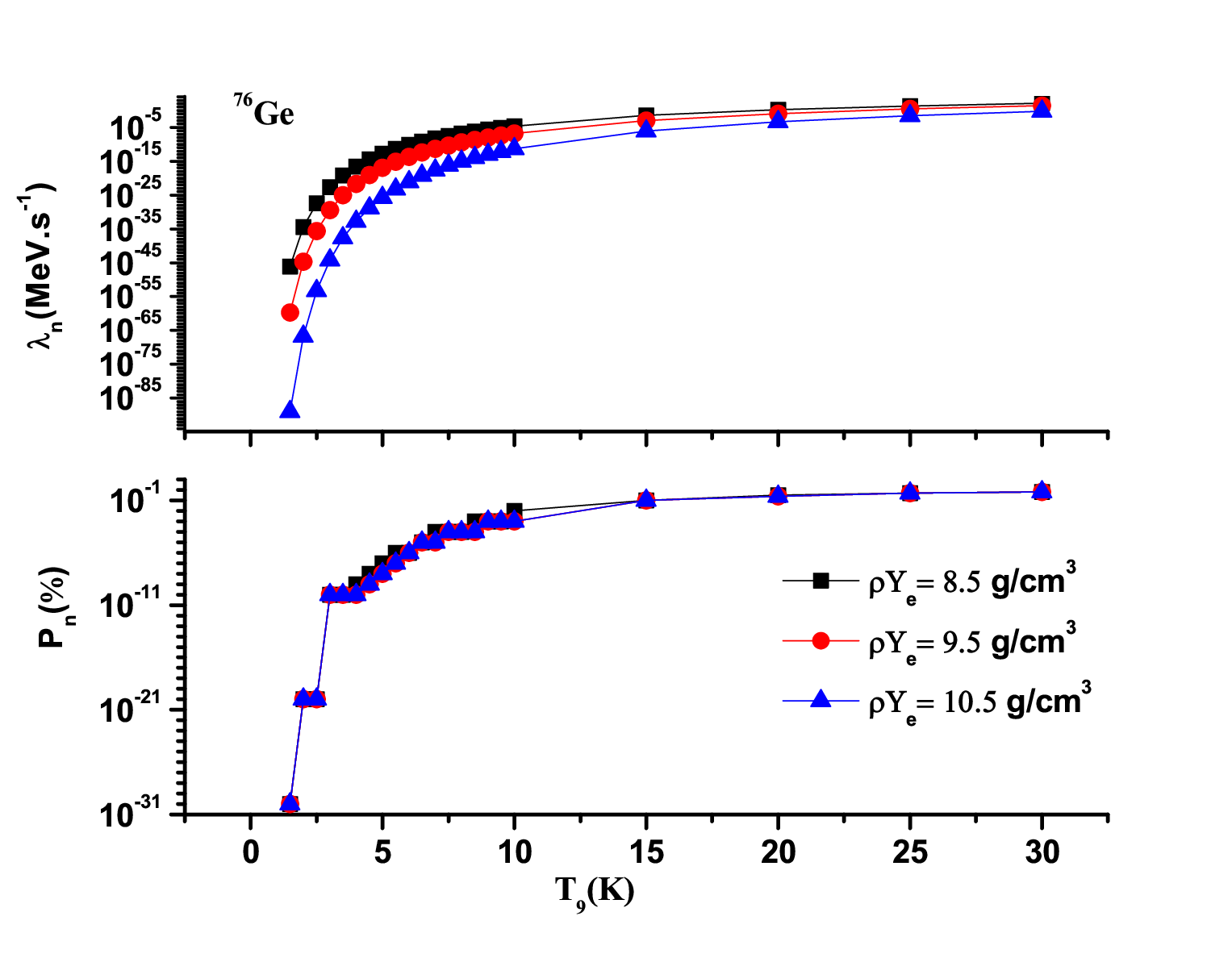}
\normalsize \caption{Energy rate of $\beta$-delayed neutron (top
panel) for $^{76}$Ge as a function of stellar temperature and
density. The bottom panel depicts the  probability of
$\beta$-delayed neutron emissions.}\label{pn76}
\end{figure}

\begin{figure}
\includegraphics[scale=0.5]{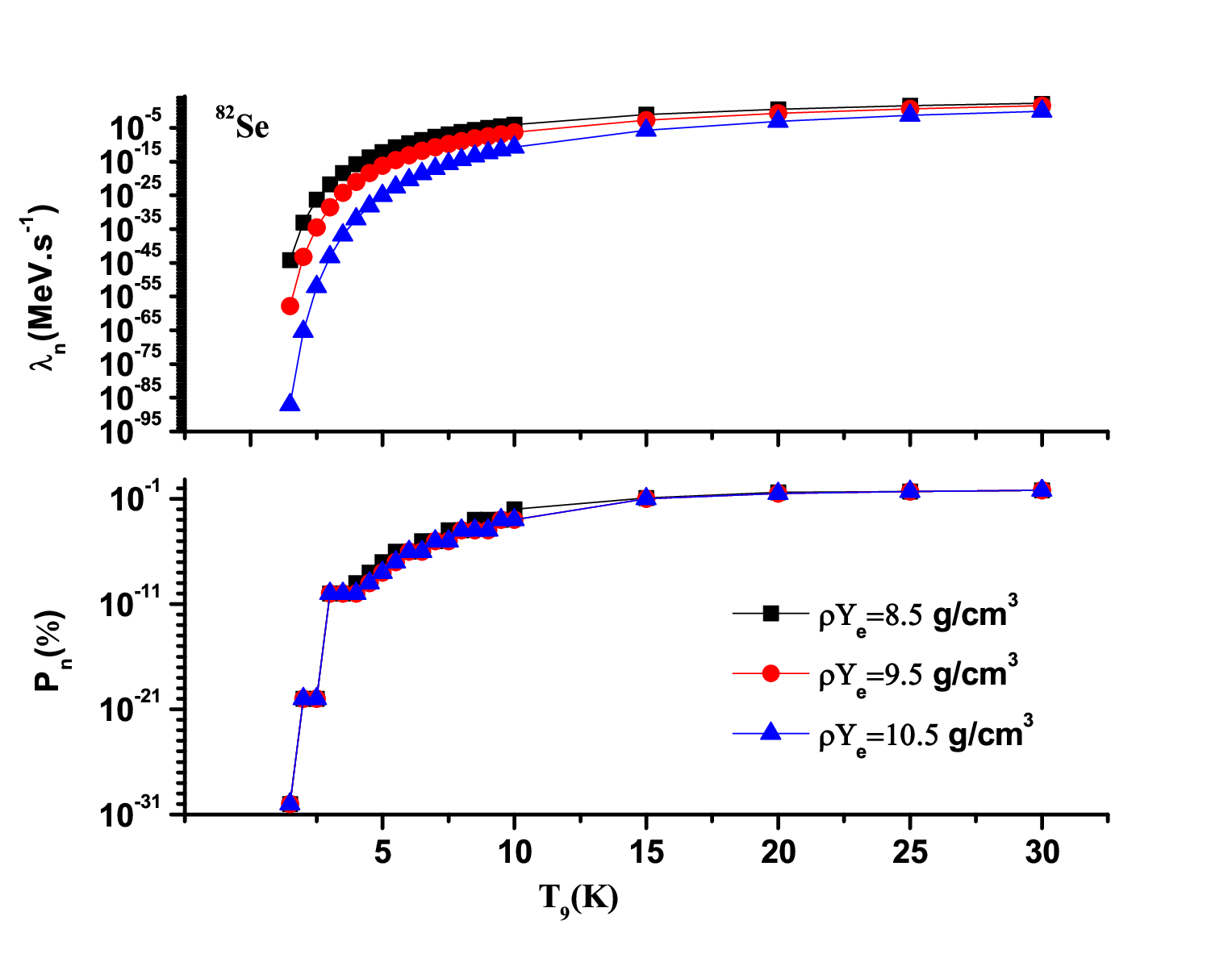}
\normalsize \caption{Same as Fig.~\ref{pn76} but for
$^{82}$Se.}\label{pn82}
\end{figure}

The energy rates of neutron emission as a function of temperature
for selected densities 8.5, 9.5 and 10.5 g $cm^{-3}$ and
corresponding probabilities of $\beta$-delayed neutron emissions are
shown in Fig.~\ref{pn76} for $^{76}$Ge and in Fig.~\ref{pn82} for
$^{82}$Se, respectively. The $\beta$-delayed neutron emission rates
are given in units of $MeV.s^{-1}$. The neutron emissions increase
with increasing stellar temperature and decreasing density. The
reason can be traced to the calculation of phase space factors which
increases with increasing temperature and decreasing density.
Likewise the probability of $\beta$-delayed neutron emissions
becomes finite at high stellar temperatures (see bottom panel of
Fig.~\ref{pn76} and Fig.~\ref{pn82}). The ASCII files of all
calculated stellar rates are available and may be requested from the
authors.

\section{Conclusions}
We calculated ground and excited state GT strength distributions for
medium-heavy nuclei, $^{76}$Ge and $^{82}$Se, using the pn-QRPA
model with deformed Nilsson basis states. These double beta decay
nuclei were selected to study the low-lying part of the GT
distribution. The low-lying GT strength functions for these $\beta
\beta$-decay nuclei may be used to provide constraints on
calculations of the matrix elements involved in the initial step of
the double beta decay process \citep{mad89}. This in turn can lead
to stringent limits on the mass of the electron neutrino via
comparison of measured and calculated lifetimes for $\beta
\beta$-decay nuclei.

The GT strength distributions of $^{76}$Ge and $^{82}$Se calculated
in the present work were found to be in decent agreement with the
measured charge-changing distributions. We later used the same model
to calculate weak interaction rates in stellar matter. These
included $\beta$-decay, positron capture, energy rates of neutron
emission and probability of $\beta$-delayed neutron emission rates.
Our calculation showed that at low stellar temperatures (T$_{9} \leq
$ 5), the positron capture rates on $^{76}$Ge and $^{82}$Se  can
easily be neglected in comparison to the competing $\beta$-decay
rates. At high stellar temperatures (T$_{9} >$ 10), the positron
capture rates command the total weak rates. The microscopically
calculated $P_n$ values, presented in this work, may influence the
final abundance of heavy elements synthesized in the r-process.

During the collapse phase allowed $\beta$-decays become unimportant
due to the increased electron chemical potential which drastically
reduces the phase space for the beta electrons. It is well known
that a smaller lepton fraction disfavors the outward propagation of
the post-bounce shock waves, as more overlying iron core has to be
photo-dissociated. The microscopic calculation of positron capture
and $\beta$-decay rates, presented in this work, may contribute to a
better understanding of the nuclear composition and $Y_{e}$ in the
core prior to collapse and collapse phase. The $\beta$-decay rates
for medium-heavy nuclei presented in this work, can assist in a more
vigorous URCA process and may lead to cooler presupernova cores
consisting of lesser neutron-rich matter than in presently assumed
simulations.

\acknowledgments  J.-U. Nabi would like to acknowledge the support
of the Higher Education Commission Pakistan through the HEC Project
No. 20-3099.

\onecolumn
\newpage

\clearpage \centering

\begin{table}
\centering \small \caption{Measured and calculated total GT strength
and centroid values for $^{76}$Ge and $^{82}$Se along $\beta$-decay
direction. For references see text.}\label{cent table}

    \begin{tabular}{c|c|c|c|c}

           Nucleus & Model &$\sum$ B(GT$_{-}$) & $\bar{E}_{-}$ (MeV) & Cut off energy(MeV)\\
\hline

           & EXP & 19.89 & 9.10 & 11.13 \\
    $^{76}$Ge & Ha \& Cheoun 2015 & 24.28 & 10.20 & 16.90 \\
           & This work & 16.30 & 8.66 & 26.50 \\

           \hline
          & EXP & 21.91 & 10.17 & 12.00  \\
    $^{82}$Se & Ha \& Cheoun 2015 & 32.46 & 11.61 & 19.19 \\
           & This work & 18.4 & 9.04 & 26.60 \\

           \end{tabular}
\end{table}

\begin{table}
\centering \small \caption{Ratio of calculated positron capture to
$\beta$-decay rates as a function of stellar temperature and
density.}\label{ratio table}
    \begin{tabular}{c|c|c|c|c|c|c}

           Nucleus & $\rho$ $\it Y_{e}$  & \multicolumn{4}{c}{$R(pc/bd)$}\\
\cline{3-7} & &T$_{9}$=01 & T$_{9}$=05 & T$_{9}$=10 & T$_{9}$=20 & T$_{9}$=30 \\
\hline

           & 8.50 & 1.28E-03 & 2.40E-02 & 1.94E+00 & 3.90E+01 & 3.45E+02 \\
    $^{76}$Ge & 9.50 & 2.22E-04 & 4.78E-02 & 4.48E+00 & 1.08E+01 & 1.47E+02 \\
           & 10.50 & 2.00E-04 & 4.19E-02 & 3.90E-01 & 8.13E+00 & 9.25E+01 \\
                     \hline
           & 8.50 & 7.52E-06 & 5.95E-03 & 1.20E-01 & 9.04E+00 & 1.79E+02 \\
    $^{82}$Se & 9.50 & 2.40E-07 & 1.47E-03 & 3.55E-02 & 2.77E+00 & 7.79E+01 \\
           & 10.50 & 1.82E-07 & 1.29E-03 & 3.18E-02 & 2.19E+00 & 5.01E+01 \\
                      \end{tabular}
\end{table}

\end{document}